# Potential Upgrade of the CMS Tracker Analog Readout Optical Links Using Bandwidth Efficient Digital Modulation

S. Dris [a, b], C. Foudas [b], K. Gill [a], R. Grabit [a], J. Troska [a], F. Vasey [a]

[a] CERN, 1211 Geneva 23, Switzerland
[b] Imperial College London, UK

Stefanos.Dris@cern.ch

*Abstract*

The potential application of advanced digital communication schemes in a future upgrade of the CMS Tracker readout optical links is currently being investigated at CERN. We show experimentally that multi-Gbit/s data rates are possible over the current 40 MSamples/s analog optical links by employing techniques similar to those used in ADSL. The concept involves using one or more digitally-modulated sinusoidal carriers in order to make efficient use of the available bandwidth.

## I. Introduction

The current CMS Tracker optical links employ analog Pulse Amplitude Modulation (PAM) at 40MS/s. The Signal to Noise Ratio (SNR) of the system is specified so that the link has an equivalent digital resolution of at least 8 bits. Hence, the analog modulation scheme is akin to digital baseband PAM using 256 distinct levels (8 bits) at 40MHz. The equivalent data rate is 320Mbits/s (=8×40MHz).

The next iteration of the CMS Tracker will be operated in the Super LHC (SLHC) environment, and will have to cope with significantly increased data rates due to the tenfold increase in luminosity that is foreseen. In contrast to the telecoms industry where the optical fiber and its installation drive the cost of a transmission system, it is the cost of the optoelectronic components that represents a large fraction of the CMS Tracker electronics budget. The high cost of development of new components able to match the physical and environmental constraints of a high energy physics experiment provides the motivation behind re-using the current link components.

It is proposed to convert these links to a digital system in order to achieve higher data rates using the current analog link components. Additional components on either side of the existing links would be required to perform the necessary digitization, digital transmission and reception. The development can be treated as that of a generic digital communication system, where the medium (or *channel*) over which digitally modulated signals are to be sent is the current CMS Tracker optical link (Figure 1). The main constraints are the bandwidth of the link and available signal power, limited by the transmitter (the Analog OptoHybrid or AOH) [1] and the 12-channel Analog Optoelectronic Receiver (ARx12 ) [2]. Therefore a bandwidth efficient digital modulation scheme −such as one based on Quadrature Amplitude Modulation (QAM)− is required to achieve transmission at Gbit/s rates. The feasibility of such a conversion must be explored in terms of performance that can be achieved and implementation complexity.

The analytical method to calculating the achievable data rate using QAM in the current Tracker optical links has been detailed in [3]. The previous work was based on simplified assumptions regarding the noise in the optical link, which was presumed to be additive white Gaussian (AWGN) with power spectral density (PSD) derived from the specifications of the readout system. At a bit error rate of $10^{-9}$, the achievable data rate was estimated at ~2.3-2.7Gbit/s [3].

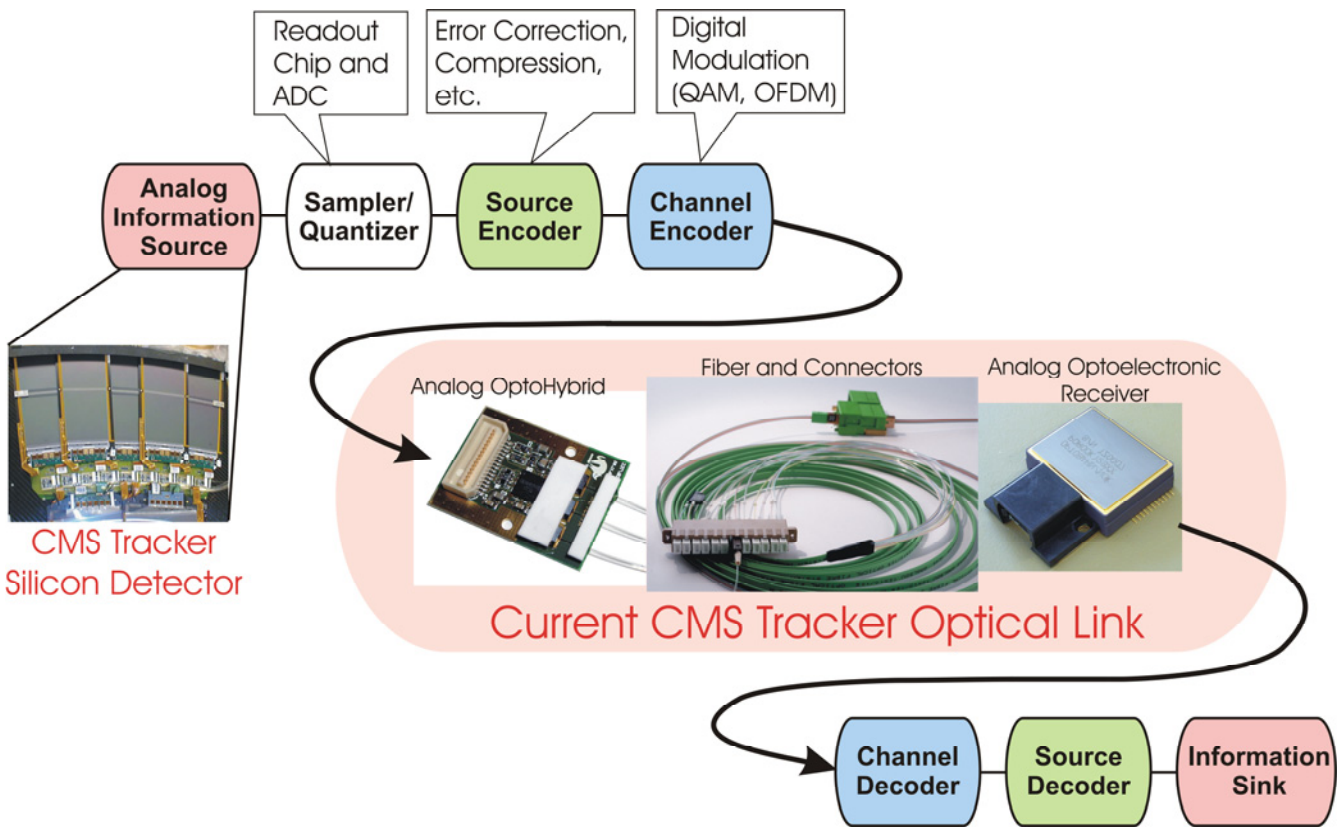

Figure 1: Basic components of a digital communication system based on the current optical link components. The analog channel is the current CMS Tracker optical link. This paper deals only with the Channel Encoder/Decoder blocks.

A future system based on QAM modulation could be implemented using a single carrier, or with multiple carriers as in Orthogonal Frequency Division Multiplexing (OFDM) and Discrete MultiTone (DMT) [4]. While there are practical limitations that will govern this decision, for analytical purposes the two systems can be considered to be equivalent in terms of data rate[1].

Experimental tests are needed, not only to test the validity of the assumptions made previously, but also to demonstrate whether or not the concept is viable and worth pursuing. Given the time constraints and complexity issues, the implementation of a custom-built prototype QAM-based digital communication system is not a realistic option at this early stage of development. Commercial QAM-based modems with such a high throughput do not exist. In order to experimentally assess whether such a system is feasible and determine the maximum theoretical data rate, a different approach is necessary.

[1] This ignores the limitations due to Peak to Average Power Ratio (PAPR) which is normally an order of magnitude higher in a typical OFDM implementation, compared to a single-carrier QAM system. More details on PAPR can be found in [4].

We consider a multi-carrier (OFDM) system with hundreds of low-symbol rate (1MS/s) QAM carriers covering the entire frequency range of the channel (~500-700MHz). The concept involves a test-setup based on commercial instruments (a signal generator and a spectrum analyzer) capable of QAM modulation/demodulation which can send and receive a single, 1MS/s QAM-modulated carrier through the optical link. If the carrier frequency is swept through the entire frequency range, each carrier of the OFDM system is effectively tested independently (Figure 2). In order to reach useful conclusions, the available transmission power needs to be appropriately allocated to each carrier, taking into account the total power constraint of the channel. Hence a multi-carrier implementation is assessed by the aggregate performance of all single-carrier tests.

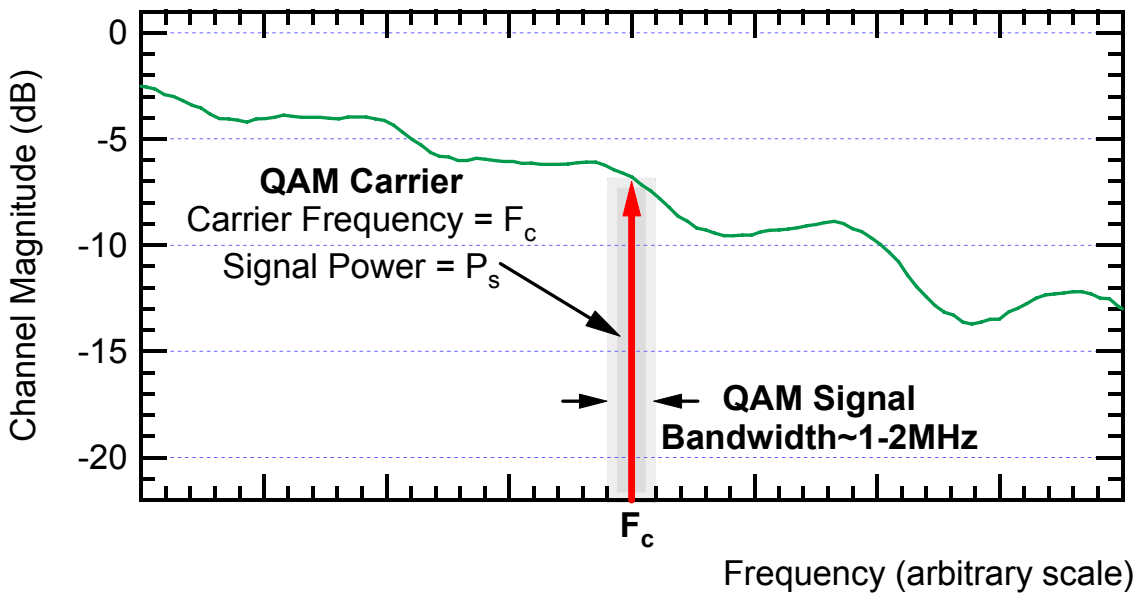

Figure 2: Illustrating a 1MS/s QAM carrier transmitted through a channel with an arbitrary frequency response. The carrier frequency, $F_c$, can be varied across the entire frequency of the channel.

## II. Testing Digital Communication Systems

### A. *QAM Signal Generation*

Combined amplitude and phase modulation is achieved by simultaneously impressing two separate *k*-bit symbols on two quadrature sinusoidal carriers (Figure 3). Each branch in Figure 3 is essentially the amplitude modulation (PAM) of the corresponding carrier.

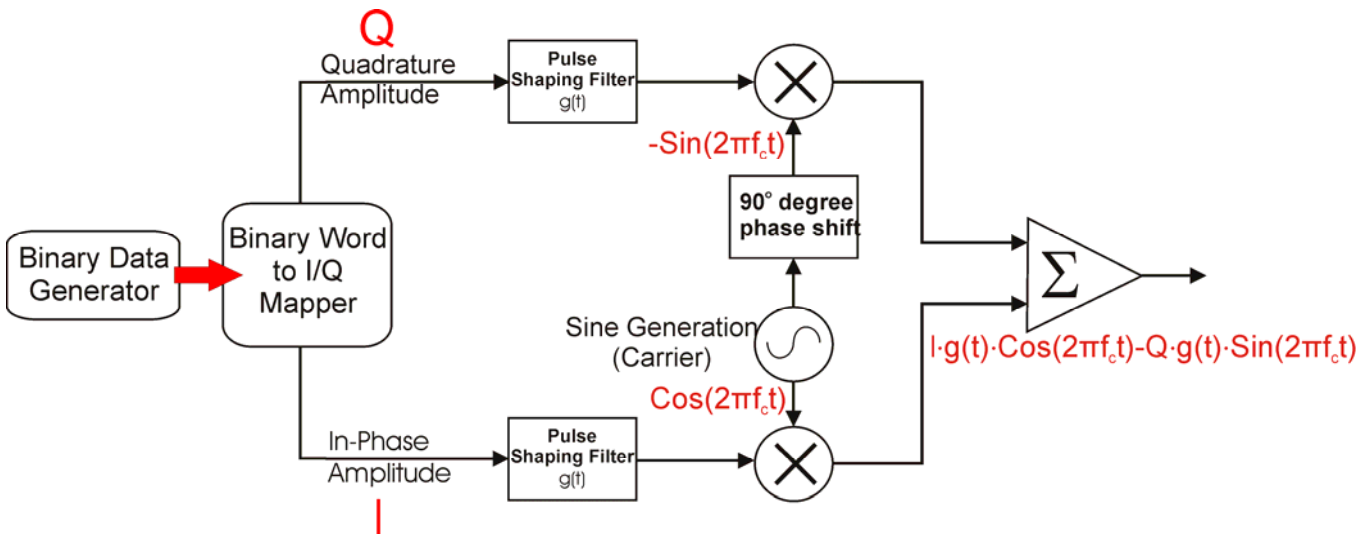

Figure 3: Illustrating QAM signal generation.

A symbol consisting of 2*k* binary bits is clocked in to the I/Q mapper at the required symbol rate. Half of the bits are translated to a quadrature amplitude, *Q*, and half to an in-phase amplitude, *I*. After pulse shaping, these are modulated on the sin and cos carriers respectively, and summed at the output. The resulting signal waveforms are:

$$\begin{aligned} s(t) &= \mathrm{Re}\left[ (I + jQ) g(t) e^{j2\pi f_c t} \right] \quad \mathit{for}\ 0 \le t \le T \\ &= I \cdot g(t)\cos(2\pi f_c t) - Q \cdot g(t)\sin(2\pi f_c t) \end{aligned} \tag{4}$$

*g(t) is the signal pulse and I and Q are the information-bearing signal amplitudes of the quadrature carriers.*

### B. *Assessing System Performance*

A basic explanation of the underlying concepts behind the experimental tests is required to aid understanding of the subsequent sections. The aim of the laboratory tests is to determine the number of bits/symbol that can be allocated to each QAM carrier [3] in a hypothetical multi-carrier system, for a given target error rate and transmission power.

While a direct BER test is the best way of evaluating a system's performance, it requires fairly complex hardware and can be very time-consuming when the target BER is low. As a guideline, for a target BER of $10^{-9}$, $3{\cdot}10^{9}$ bits need to be transmitted through the link with no errors for a 95% confidence level in the result. The tests to be carried out will involve individual, low symbol rate carriers. For example, a 1MS/s carrier with 5bits/symbol transmits 5Mbits/s, requiring 10 minutes to complete the test for an error rate of $10^{-9}$. This is unacceptably slow, given that the tests will involve sweeping the carrier frequency, input power and bit assignment, hence requiring hundreds of individual BER tests. A more practical approach is described next.

### C. *Error Vector Magnitude*

Error Vector Magnitude (EVM) is a modulation quality metric widely used in digital RF communications [5]. It is sensitive to any impairments that affect the amplitude or phase of a demodulated signal. The error vector is defined as the vector difference between the reference signal (i.e. the ideal constellation points) and the measured (demodulated) signal in the IQ plane[2] (Figure 4). EVM is the root mean square (rms) error vector over time, and is usually normalized, to the *average* symbol magnitude of the respective QAM constellation. By convention, it is expressed as a percentage:

$$\mathrm{EVM} = \frac{\text{Error Vector (rms)}}{\text{Average QAM Signal Amplitude}} \cdot 100\% \tag{1}$$

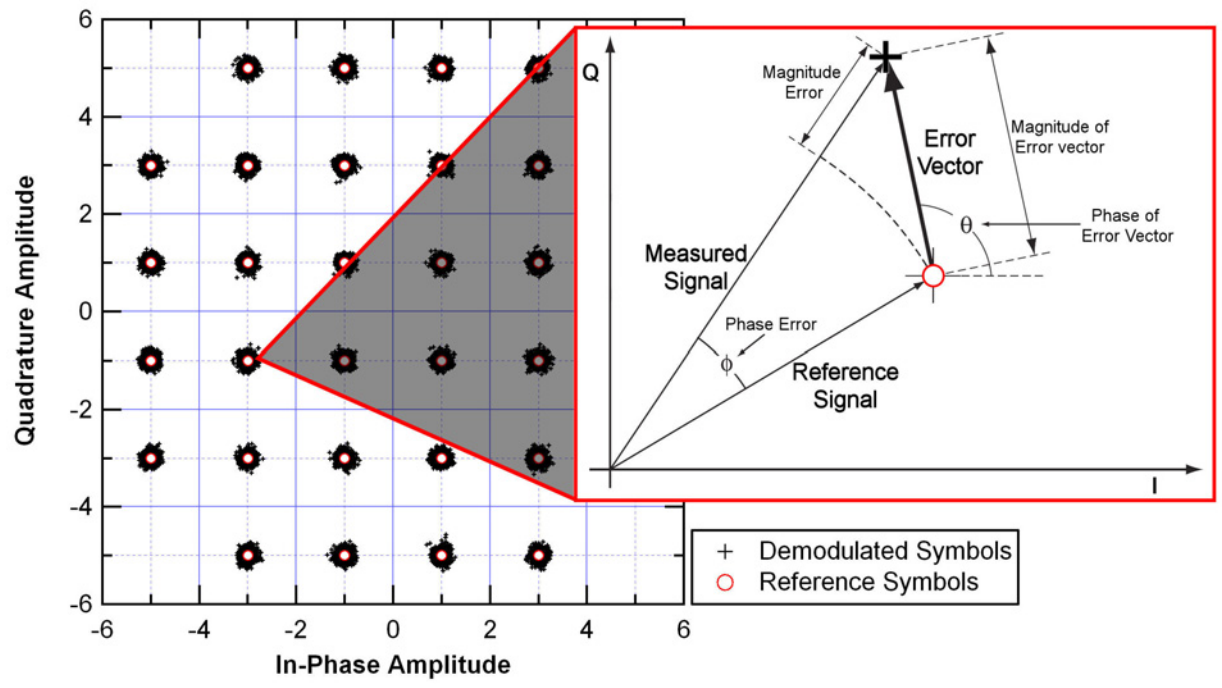

Figure 4: Error vector and related quantities. The IQ plot shows 40 000 demodulated symbols belonging to a 32-QAM constellation. The EVM is ~3.9% in this case, corresponding to an SNR of 28.2dB.

---

[2] An IQ plot is a useful way of representing demodulated QAM signals. The in-phase Vs quadrature amplitudes of the QAM signal are plotted. More details on this can be found in [3, 4].

EVM can be related to the digital Signal to Noise Ratio (SNR) by:

$$\mathrm{EVM} = \frac{1}{\sqrt{\mathrm{SNR}}} \qquad (2)$$

### *D. Symbol Error Rate from EVM*

It has been shown how the SNR in a QAM digital communication system can be derived from an EVM measurement. The significance of this is that SNR can be used to relate EVM to the error rate. For a system corrupted by Additive White Gaussian Noise (AWGN) the symbol error rate[3] (SER) in QAM is given by [6]:

$$SER \approx 2erfc\left(\sqrt{\frac{3 \cdot SNR}{2 \cdot (2^b - 1)}}\right) \qquad (3)$$

*Where erfc is the complementary error function and b the number of bits per symbol (e.g. 4 for 16-QAM).*

Equation (3) is graphically depicted in Figure 5 for various modulation schemes, where the probability of error is plotted against SNR. Figure 6 shows the corresponding relationship with EVM.

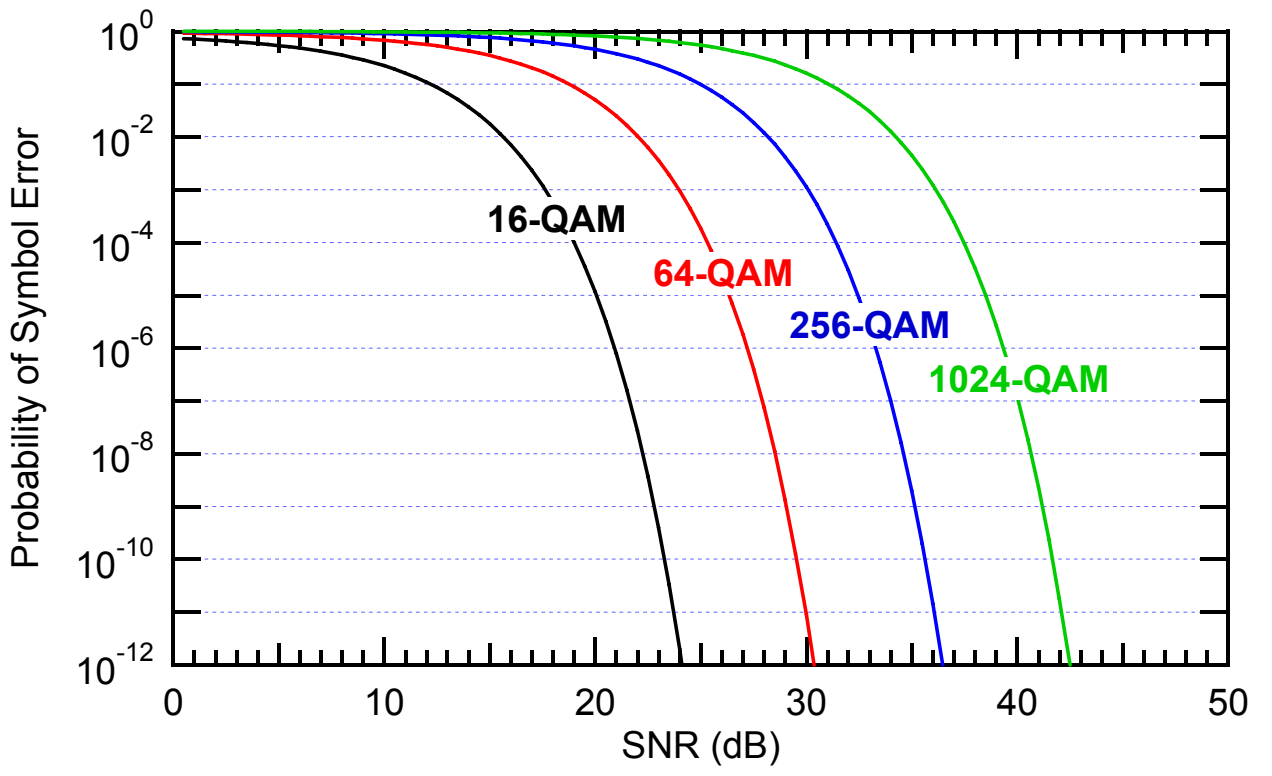

Figure 5: Probability of symbol error in QAM as a function of SNR.

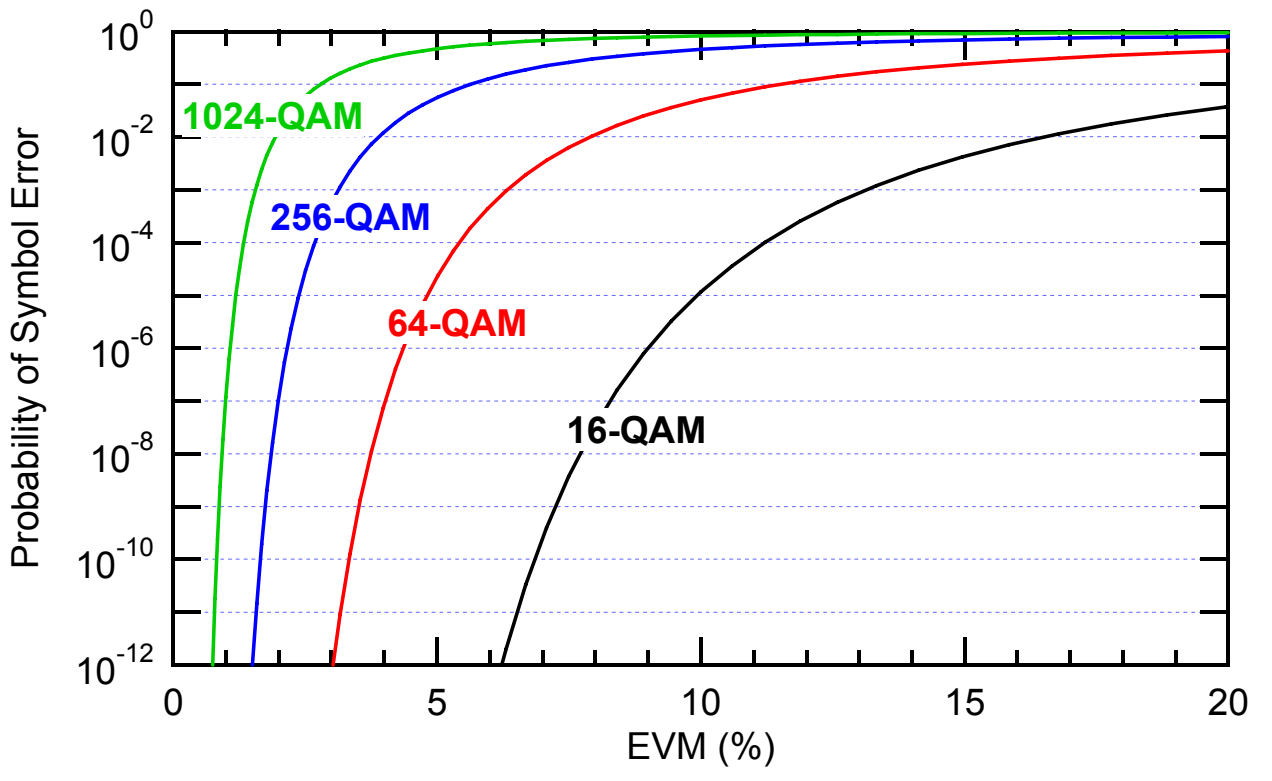

Figure 6: Probability of symbol error in QAM as a function of EVM.

[3] The relationship between bit error rate (BER) and SER depends on the bit-mapping and the QAM constellation. The BER is upper bounded by the SER multiplied by the number of bits/symbol, *b*. For rectangular Gray-encoded QAM constellations where the SER>>1, $\mathrm{BER} \approx \mathrm{SER}/b$. This discussion is beyond the scope of this paper, and we will limit ourselves to using the SER.

The advantages of using EVM measurements to estimate the error rate, rather than a direct BER test, are that of speed and lower complexity. There is no need to decode the demodulated symbols. EVM can be accurately estimated using only a few thousand symbols, in contrast to a BER test that requires sending and receiving a large number of bits to allow an accurate calculation.

## III. Laboratory Tests

### *A. Objectives*

The aim of the tests described in this paper is to determine, experimentally, the data rate achievable using bandwidth efficient digital modulation in a link based on the current optical link components. To achieve this, it is necessary to determine the number of bits/symbol that can be allocated to each of the QAM carriers across the entire frequency range of the link, given the total available transmission power and the target error rate. Since it is the SNR of a QAM carrier that determines the number of bits/symbol that can be allocated to it (for a given target error rate), it follows that the SNR as a function of carrier frequency and transmission power is all the information needed to calculate the data rate.

### *B. Test Setup*

An Agilent ESG E4438C vector signal generator [7] was used as a modulator, while an Agilent PSA 4440A spectrum analyzer [8] demodulated the received QAM signals. The optical link consisted of production versions of the AOH and ARx12. One AOH channel's fiber pigtail was mated to one of the connectors on a 12-channel MU/MPO12 patch cord, with the MPO connector mated to the ARx12. The AOH requires a differential input, while the signal generator has a 50Ω single-ended RF output. An interface based on the high-bandwidth AD8351 differential driver was therefore employed. On the receiving end, the spectrum analyzer was operated in ac-coupled input mode, receiving the single-ended output from the ARx12 directly. 50Ω coaxial cables with type-N and BNC connectors were used for the electrical connections between the instruments and components. A Mac computer running Labview and equipped with GPIB provided control for the AOH and ARx12, with an Agilent 34970A switch used to select the appropriate settings on the optical receiver. Finally, a GPIB-equipped PC was used for communication with the signal generator and spectrum analyzer. The test setup is shown schematically in Figure 7.

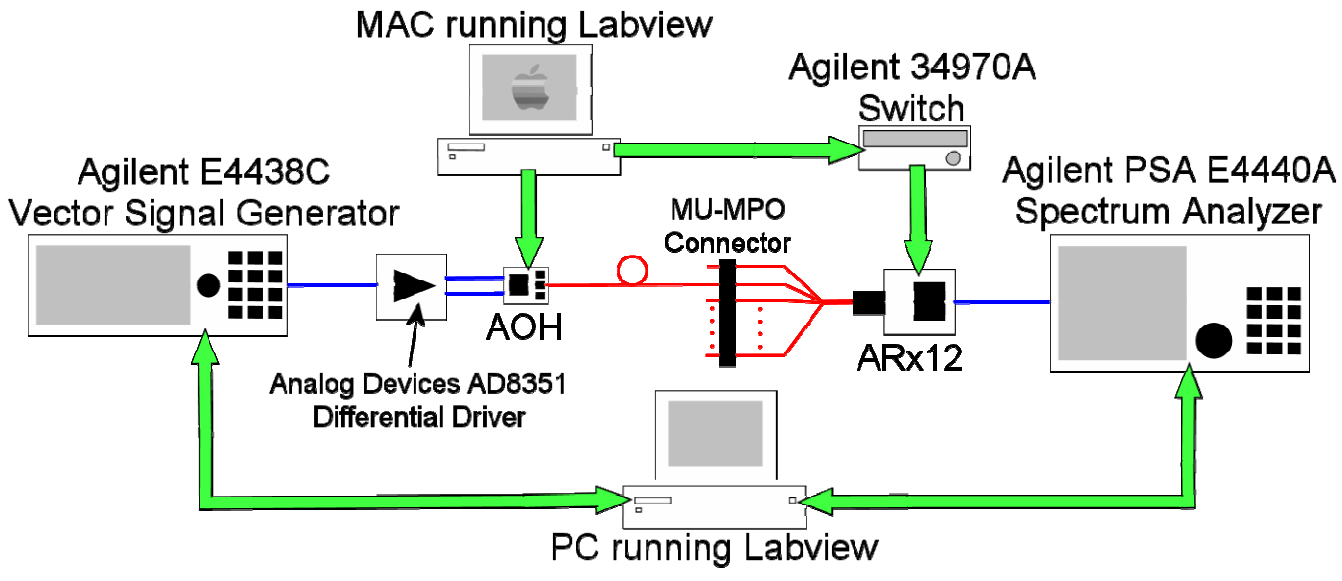

Figure 7: Showing the test setup used. Control and data acquisition (indicated by the arrows) was performed using GPIB-equipped computers.

### C. Method

Five different modulation schemes were used in the tests: 4, 16, 32, 64 and 256-QAM. The modulated carrier was swept in frequency (20MHz to 1000MHz, in 20MHz steps) and input power[4] (0dBm to ~ -50dBm), for each modulation scheme (Figure 8). The symbol rate was kept at a constant 1MS/s, and root Nyquist pulse shaping with a filter alpha of 1 was used at the transmitter and the receiver. The internal baseband generator of the vector signal generator was set to produce a pseudo-random bit pattern which was then used to modulate the carrier. For every point in frequency and power, the average EVM and SNR was calculated over 2 000 received symbols.

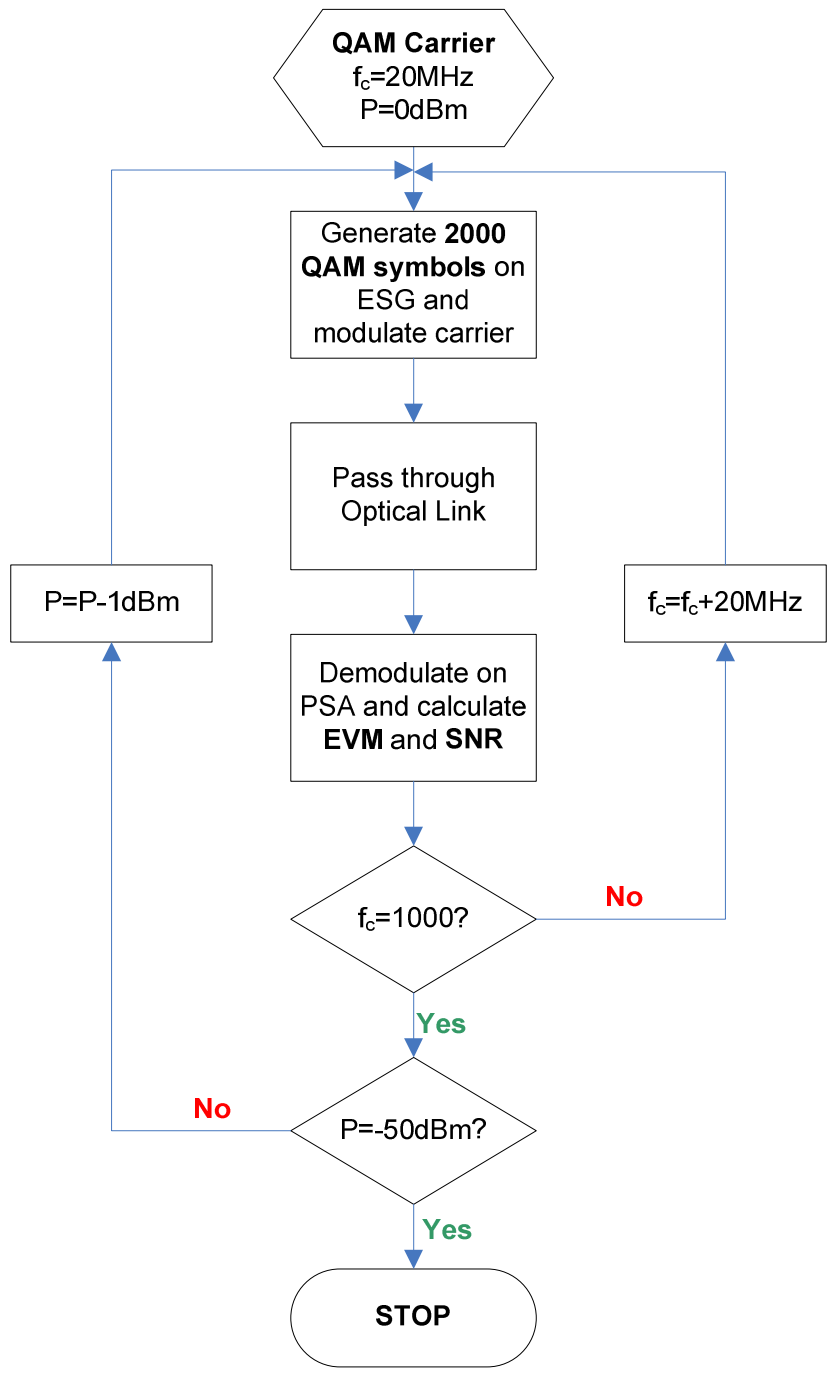

Figure 8: QAM test flowchart, showing the procedure followed for each modulation scheme (4-QAM, 16-QAM, 32-QAM, 64-QAM, 256-QAM).

### D. Results

#### 1) Noise

If the SNR and error rate of a given QAM carrier are to be calculated from the EVM, it is important to determine whether or not the noise in the test system is indeed Gaussian (since equations (2) and (3) are only valid for AWGN channels). Several IQ plots were taken at various frequencies and input powers. The results obtained were similar for every case where there was no amplifier saturation in the link. A typical case is presented in this section.

A 32-QAM signal with a symbol rate of 1MS/s and a carrier frequency of 40MHz was passed through the link (see Figure 4 for the noisy, received constellation). 40 000 symbols were retrieved by the spectrum analyzer. In order to obtain maximum statistics on the noise, the received IQ points were amplitude normalized relative to their ideal symbol points and merged into one. Moreover, in the subsequent tests, the EVM calculated is an average over all received symbols, not one particular constellation point. Hence, studying the noise in this way is consistent with the test method.

A 2-dimensional histogram of the received (noisy) points is shown in Figure 9. The projections of the in-phase and quadrature components of the histogram are clearly Gaussian, and hence the EVM can be used to obtain the SNR (and error rate) in a QAM system, with a high degree of accuracy.

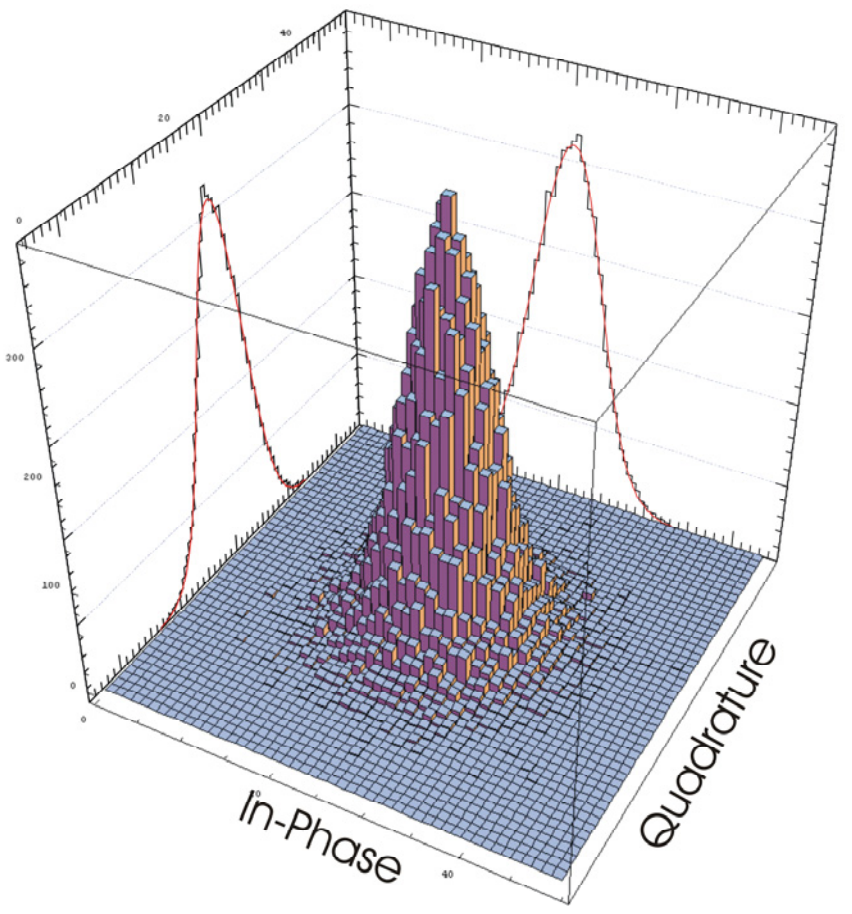

Figure 9: 3-D representation of the 2-D histogram of the 40 000 received IQ points. Projections of the 1-D in-phase and quadrature histograms are also shown.

#### 2) SNR Vs Carrier Frequency

The same test procedure (Figure 8) was followed for all modulation schemes (4, 16, 32, 64 and 256-QAM). Hence, for a range of input powers, plots of SNR Vs carrier frequency were obtained. The results were combined into one graph by averaging the SNRs obtained at each frequency and for each modulation scheme (Figure 10). By sweeping the frequency and input power, the parameter space of the two quantities which are needed for the link's capacity calculation have been fully explored. The error bars in Figure 10 denote the standard deviation of the SNR values. The results were quite consistent across all modulation schemes, giving confidence in the test method [9].

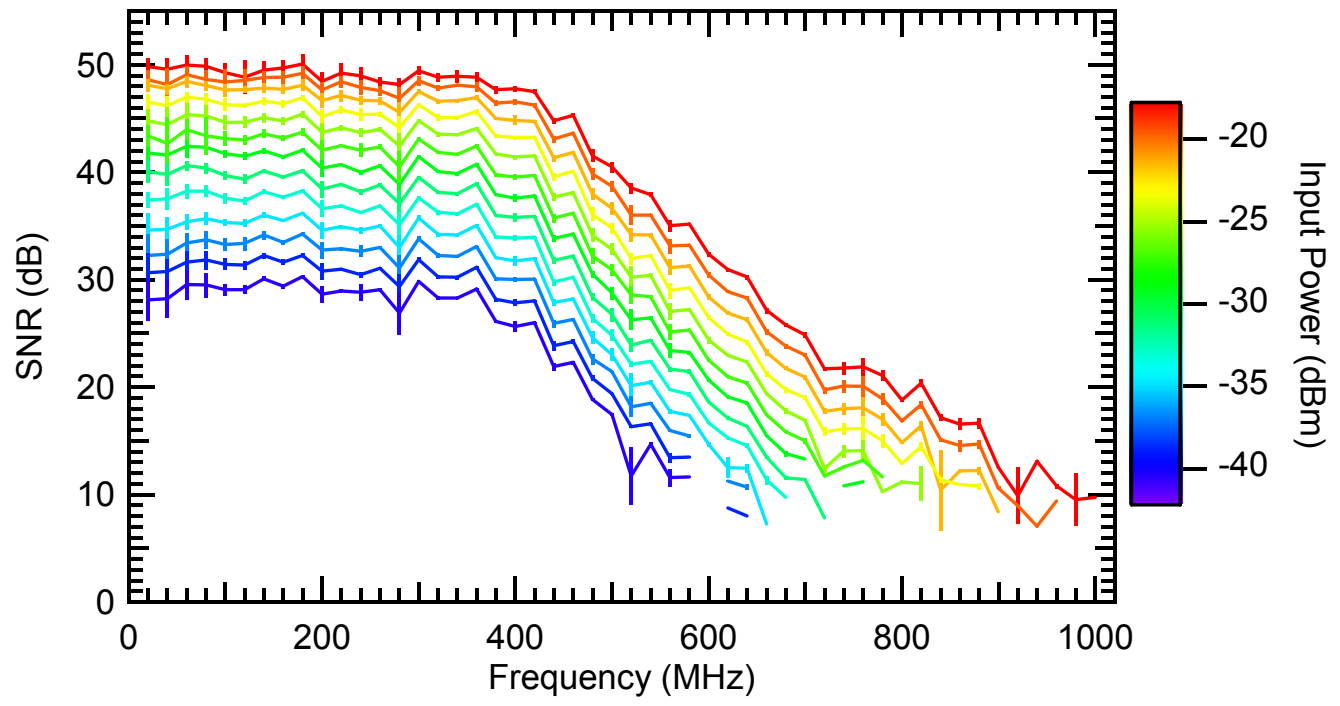

Figure 10: SNR Vs frequency for a carrier of symbol rate 1MS/s, for several input powers. The plots were obtained by averaging the results from all tests (4, 16, 32, 64, 256-QAM).

[4] The input power stated in this paper is that reported by the signal generator, and is not necessarily equal to the electrical transmission power injected at the input of the optical link. The two quantities are, however, proportional.

## IV. ACHIEVABLE DATA RATE

In the previous section, the SNR Vs center frequency of a QAM carrier was calculated for a range input powers. This gives all the information required to calculate the data rate of a QAM-based system over the optical link. The analytical method described in [3, 9] is used for the calculation. It should be noted that the amount of transmission power available for such a system is implementation-specific. Therefore, the data rate is calculated as a function of normalized transmission power[5], for various target error rates, and is shown in Figure 11. The shaded area indicates the likely operating range of a future QAM system (whether single or multi-carrier), leading to predicted data rates of 3.5-4.0 Gbit/s for a target symbol error rate of $10^{-9}$.

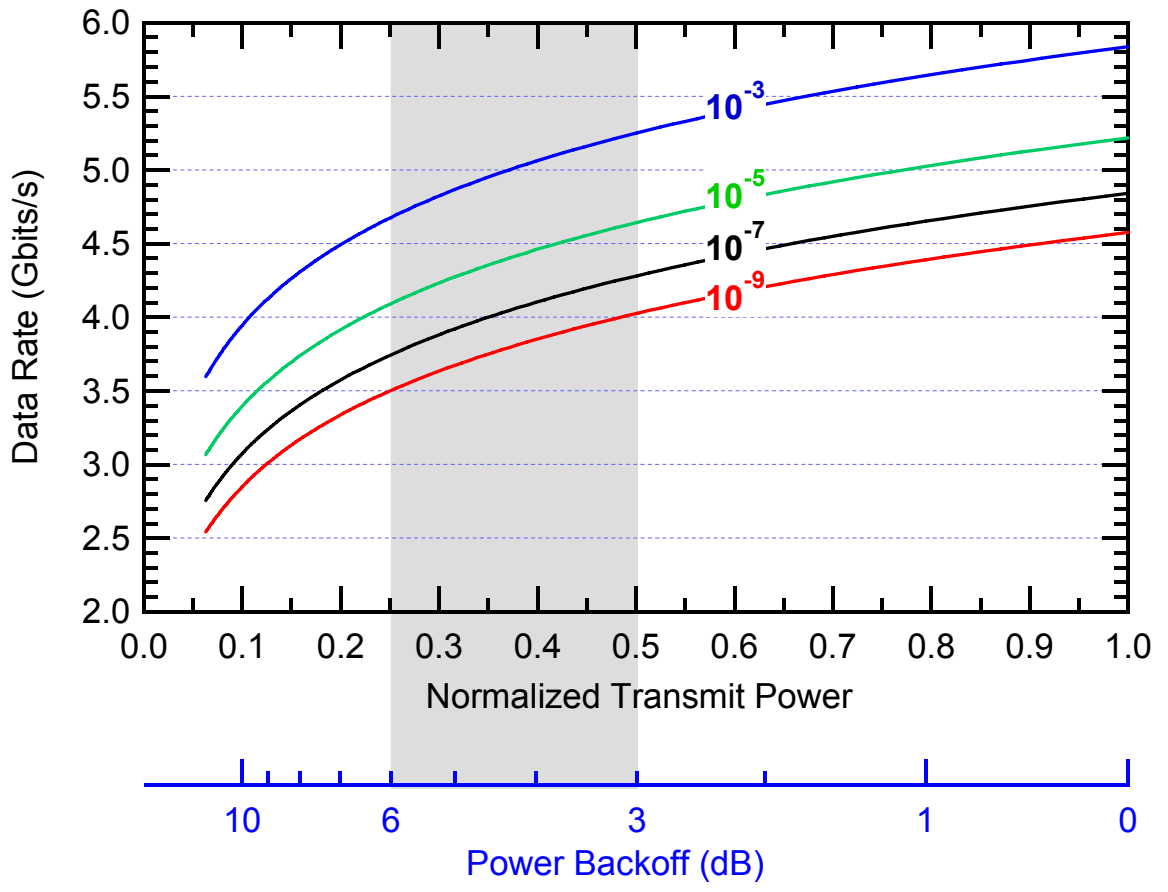

Figure 11: Data rate as a function of transmission power, for various target error rates.

## V. CONCLUSIONS

It has been demonstrated experimentally that ~3.5-4.0Gbit/s at an error rate of $10^{-9}$ can be achieved using uncoded QAM-based digital modulation over the CMS Tracker analog optical links. Compared to the data rate of 2.3-2.7Gbit/s calculated in the analytical approach followed in [3], the results are much more optimistic. The main reason for this is that the assumptions made previously turned out to be highly conservative. The noise specification of the link was used in the calculation, and assumed to be additive white. Furthermore, the system was assumed to be a frequency selective channel with a constant noise source at the output. This is clearly an over-simplified approach, which greatly affected the SNR calculation. Finally, due to the better than expected SNR, the available channel bandwidth was larger by about 150-200MHz, with a corresponding increase in overall data rate.

Whether the estimated data rate will be enough for a future CMS Tracker operating in the SLHC environment remains to be seen. Factors such as implementation complexity, power consumption, and detector data volume will all drive any future decision.

A QAM-based system with such performance would be unprecedented in commercial applications, meaning a future implementation would consist of proprietary hardware. This represents a departure from the conventional approach for high energy physics (HEP) applications, where industrial trends drive component and system design choices. Consequently, more effort is required for research and development, and the component costs will certainly be higher than less complex, commercially available optical systems are capable of over 10Gbit/s, and may present a more cost-effective solution.

A QAM modulator would require high speed digital signal processing and digital to analog conversion in the front-end. The power consumption of such devices may be prohibitively large for the context of a HEP experiment.

Despite the drawbacks, the proposed scheme would have the advantage of re-using an existing system, hence preserving the investment in components as well as in non-material costs (e.g. research and development, quality assurance, production testing, installation, etc.).

## VI. ACKNOWLEDGEMENTS

The authors are grateful to Jean-Jacques Gratier and Agilent Technologies for kindly lending us the instruments needed to complete these tests, and for their technical support. We also thank Christophe Sigaud for his help with the interfacing required in the test setup.

[5] Or equivalently, this can be expressed as power 'backoff' from the maximum signal power allowed by the linearity of the optical link.